\documentclass[aps,english,showpacs,twocolumn]{revtex4-1}
\usepackage{amssymb}
\usepackage{amsmath}
\usepackage{graphicx}
\usepackage{epsfig}

\begin{document}

\title{Probability-preserving evolution in a non-Hermitian two-band
model}
\author{W. H. Hu, L. Jin, Y. Li, and Z. Song}
\email[E-mail: ]{songtc@nankai.edu.cn}
\affiliation{School of Physics, Nankai University, Tianjin 300071,
China}

\begin{abstract}
A non-Hermitian $\mathcal{PT}$-symmetric system can have full real
spectrum but does not ensure probability preserving time evolution,
in contrast to that of a Hermitian system. We present a
non-Hermitian two-band model, which is comprised of dimerized
hopping terms and staggered imaginary on-site potentials, and study
the dynamics in the exact $\mathcal{PT}$-symmetric phase based on
the exact solution. It is shown that an initial state, which does
not involve two equal-momentum-vector eigenstates in different
bands, obeys perfectly probability-preserving time evolution in
terms of the Dirac inner product. Beyond this constriction, the
quasi-Hermitian dynamical behaviors, such as non-spreading
propagation and fractional revival of a Gaussian wave packet, are
also observed.
\end{abstract}

\pacs{11.30.Er, 03.65.-w, 03.75.-b}


\maketitle

\section{Introduction}
\label{sec.intro}

Hermiticity of the Hamiltonian as the fundamental postulate in quantum
mechanics guarantees the real eigenvalues and the conservation of
probability. However, the recent discovery of Bender and Boettcher
showed that Hermiticity of the Hamiltonian is not essential for a real
spectrum \cite{bender}. It has been proved that a non-Hermitian
$\mathcal{PT}$-symmetric Hamiltonian can have real spectrum
\cite{AM43,mosta1,mosta2}. Based on a time-independent inner product
with a positive-definite norm, a new class of complex quantum theories
having positive probabilities and unitary time evolution is
established. The Hermitian and the non-Hermitian Hamiltonians seem to
describe two parallel worlds, and much effort has been devoted to the
connection between them \cite{bender,AM43,mosta1,mosta2,Ahmed,Berry,%
Heiss,Jones,Muga,Bender02,BenderRPP,Tateo,AMIJGMMP,MZnojil,Bender03,%
Bender04}.

One of the ways of connecting a pseudo-Hermitian Hamiltonian with its
Hermitian counterpart is the metric-operator theory outlined in
\cite{mosta2}, providing a mapping between them. However, the obtained
equivalent Hermitian Hamiltonian is usually quite complicated
\cite{mosta2,jin}. Alternative ways, such as the interpretation of the
non-Hermitian systems in the frameworks of scattering and quantum
phase transition, have been investigated \cite{jin2,JLPRA83,JLJPA44,%
GLG2,MoiseyevBook}.

In this work, we investigate the dynamics of a
$\mathcal{PT}$-symmetric pseudo-Hermitian Hamiltonian in the context
of unbroken $\mathcal{PT}$ symmetry. We consider an exactly solvable
non-Hermitian $\mathcal{PT}$ model. It is a two-band tight-binding
ring, with the non-Hermiticity arising from staggered imaginary
potentials. It has been shown that such potentials can be realized in
the realm of optics through a judicious inclusion of index guiding and
gain/loss regions \cite{JPA38.171,PRL103904,PRL030402,OL2632}.
Recently, it was reported that the most salient character of the
pseudo-Hermitian Hamiltonian, which is the $\mathcal{PT}$ symmetry
breaking, was observed experimentally \cite{AGuo,Ruter}. Nevertheless,
the reality of the spectrum is not the unique common feature for the
pseudo-Hermitian and the Hermitian systems in some cases. In
Ref.~\cite{LJin2012} it is pointed out that some non-Hermitian
scattering centers, which consist of two Hermitian clusters with
anti-Hermitian couplings between them, can act as Hermitian scattering
centers, i.e. the S-matrix is unitary, or the Dirac probability
current is conserved. The goal of the present work is to show the
dynamical similarity between a non-Hermitian system and a Hermitian
one. Intuitively, closely localized gain and loss potentials may be
balanced with each other, or equivalently, the temporal and spatial
large-scale dynamics should be probability preserving. The Dirac inner
product can be measured in an universal manner in experiments, hence
it is of central importance to most practical physical problems. In
this work we aim at investigating the dynamical behavior in terms of
the Dirac inner product. Within the unbroken $\mathcal{PT}$-symmetric
region, the eigenfunctions with different $k$ are orthogonal
spontaneously in terms of the Dirac inner product. This feature
ensures the probability-preserving evolution of a state, which
involves only one or two subbands with different $k$. In this sense,
the non-Hermitian Hamiltonian acts as a Hermitian one without
employing the biorthogonal inner product. We also provide some
illustrative simulations to show the occurrence of the fractional
revivals and the slowly spreading of a wave packet. It shows that for
certain special models, the non-Hermitian and Hermitian Hamiltonians
can describe the same physics within a certain energy range.

This paper is organized as follows. In Sec.~\ref{sec.model}, we
present a non-Hermitian $\mathcal{PT}$-symmetric two-band model and
its exact solution. In Sec.~\ref{sec.HerCou}, we investigate the
Hermitian counterpart of this model. In Sec.~\ref{sec.quasi}, we show
the quasi-canonical commutation relations and the quasi-Hermitian
dynamics. In Sec.~\ref{sec.EffRing}, we demonstrate the results for
the system approaching to the exceptional point. Sec.~\ref{sec.discus}
is the summary and discussion.

\section{Model and solutions}
\label{sec.model}

\begin{figure}[tbp]
\includegraphics[bb=148 185 420 600,width=0.25\textwidth,clip]%
{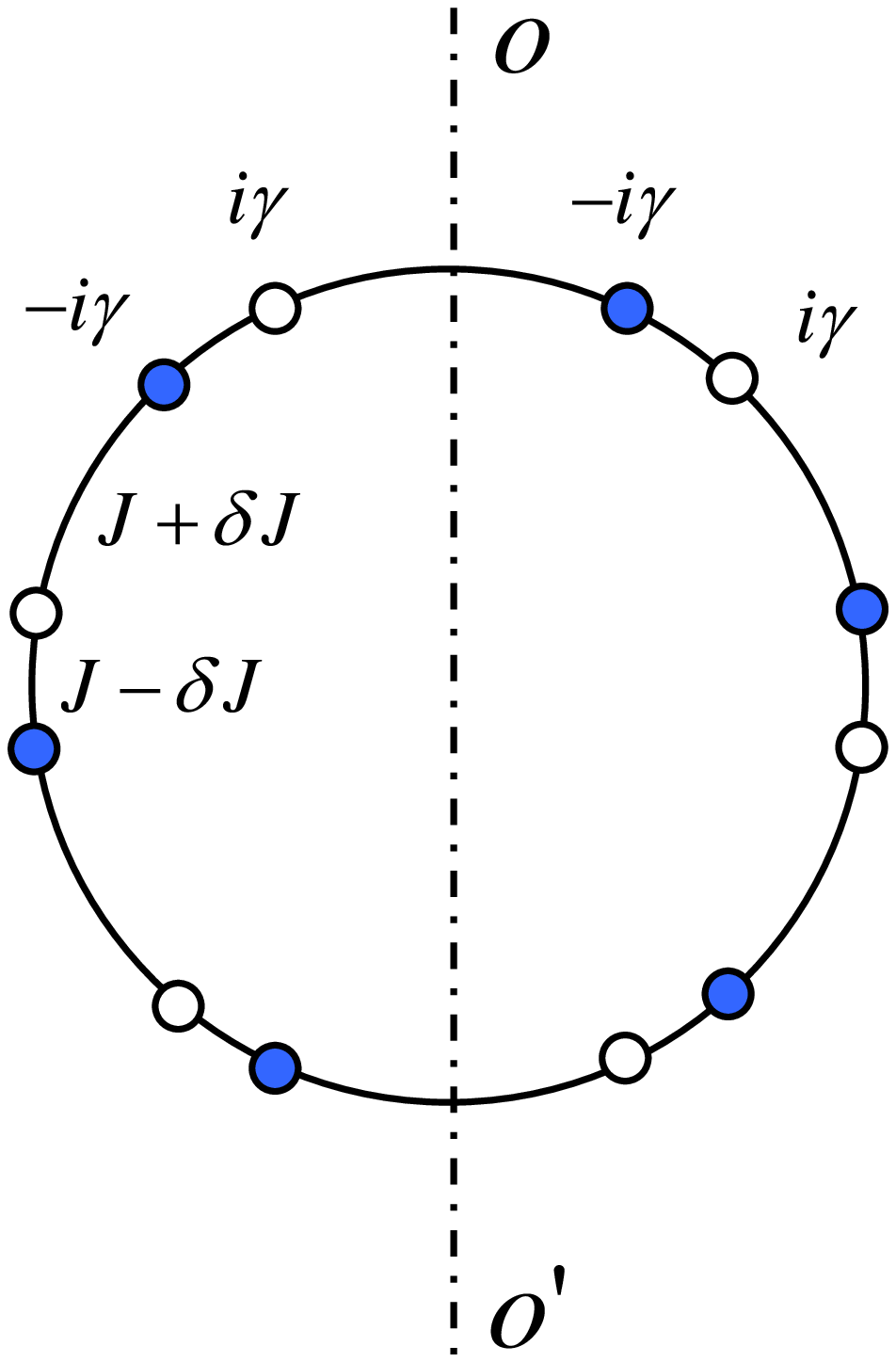}
\caption{(Color online) Schematic illustration of the Peierls
tight-binging ring with staggered imaginary potentials (solid and
empty circles). It is $\mathcal{PT}$-symmetric with respect to the
$OO^{\prime}$ axis and invariant under the translation.}
\label{illus}
\end{figure}

We consider a two-band model described by a non-Hermitian Hamiltonian
$H$. It is a tight-binding ring with the Peierls distortions between
nearest-neighboring sites and the additional staggered imaginary
on-site potentials, which can be written as follows
\begin{eqnarray}
H &=& -J \sum_{l=1}^{2N} \left[ 1+\left( -1\right) ^{l}\delta \right]
\left( a_{l}^{\dagger} a_{l+1} + \mathrm{H.c.} \right)  \nonumber \\
&& + i \gamma \sum_{l} \left( -1 \right) ^{l} a_{l}^{\dagger} a_{l},
\label{Ham0}
\end{eqnarray}
where $a_{l}^{\dagger}$ is the creation operator of a boson (or a
fermion) at the $l$th site, with the periodic boundary condition
$a_{2N+1}=a_{1}$. The hopping strengths, the distortion factor and the
alternating imaginary potential magnitude are denoted by $J$, $\delta$
and $\gamma$ ($\gamma >0$), respectively. A sketch of the lattice is
shown in Fig.~\ref{illus}. In the absence of the staggered potentials
or the Peierls distortion (with real potentials), it is a standard
two-band model and is employed to be a gapped data bus for quantum
state transfer \cite{YSJCP,HMXEPL,ChenB}. It is a
$\mathcal{PT}$-symmetric model with respect to an arbitrary diameter
axis. Here, without loss of generality, we define the action of time
reversal and parity in such a ring system as follows. While the time
reversal operation $\mathcal{T}$ is such that
$\mathcal{T} i \mathcal{T} = - i $, the effect of the parity is such
that $\mathcal{P}a_{l}^{\dagger}\mathcal{P}=a_{2N+1-l}^{\dagger}$.
Applying operators $\mathcal{P}$ and $\mathcal{T}$ on the Hamiltonian
(\ref{Ham0}), one has $\left[ \mathcal{T},H\right] \neq 0$ and
$\left[\mathcal{P},H\right] \neq 0$, but $\left[ \mathcal{PT} , H
\right] =0 $. According to the non-Hermitian quantum theory, such a
Hamiltonian may have fully real spectrum when appropriate parameters
are taken. In the following, we will diagonalize this Hamiltonian and
show that it has fully real spectrum.

Beyond the $\mathcal{PT}$ symmetry, $H$ is invariant under the
translational transformation. Then taking the Fourier transform
\begin{eqnarray}
A_{k} &=&\frac{1}{\sqrt{N}}\sum_{l=1}^{N} e ^{- i kl}a_{2l-1} ,
\nonumber \\
B_{k} &=&\frac{1}{\sqrt{N}}\sum_{l=1}^{N} e ^{- i kl}a_{2l} ,
\label{A_k}
\end{eqnarray}
where $k=2\pi n/N$, $n \in \left[ 0,N-1\right] $ is the momentum, the
original Hamiltonian can be expressed as
\begin{equation}
H = \sum_{k} H_{k}  \label{H_k}
\end{equation}
with
\begin{eqnarray*}
H_{k} &=&-J\left[ \left( 1-d+\left( 1+d\right) e ^{- i k}\right)
A_{k}^{\dagger}B_{k}+\mathrm{H.c.} \right] \\
&&- i \gamma \left( A_{k}^{\dagger}A_{k}-B_{k}^{\dagger}B_{k}\right) .
\end{eqnarray*}
Here $A_{k}^{\dagger}$ and $B_{k}^{\dagger}$ are two kinds of creation
operators of bosons (or fermions), resulting $\left[ H_{k} ,
H_{k^{\prime }} \right] =0$. The operator $H_{k}$ is non-Hermitian and
can be readily written as
\begin{equation}
H_{k} = -\epsilon_{k} \left( \bar{\alpha}_{k} \alpha_{k} -
\bar{\beta}_{k} \beta_{k} \right) ,  \label{Dia H_k}
\end{equation}
by applying the linear transformation
\begin{eqnarray}
\alpha _{k} &=& \mu_{k} A_{k} + \nu_{k} B_{k},  \nonumber \\
\beta _{k} &=& - \bar{\nu}_{k} A_{k} + \bar{\mu}_{k} B_{k} ,
\label{apha_k}
\end{eqnarray}
and
\begin{eqnarray}
\bar{\alpha}_{k} &=& \bar{\mu}_{k} A_{k}^{\dagger } + \bar{\nu}_{k}
B_{k}^{\dagger} ,  \nonumber \\
\bar{\beta}_{k} &=& - \nu_{k} A_{k}^{\dagger} + \mu_{k}
B_{k}^{\dagger} ,
\label{apha_k bar}
\end{eqnarray}
where the spectrum is given by
\begin{equation}
\epsilon _{k}=2J\sqrt{\left( 1-\delta ^{2}\right) \cos ^{2} \left(
\frac{k}{2} \right) + \delta ^{2} - \left( \frac{\gamma }{2J}\right)
^{2}},
\label{spectrum}
\end{equation}
and
\begin{eqnarray}
\mu_{k} = & \cos \theta_{k} e ^{ i \frac{\phi_{k}}{2}} , \
\bar{\mu}_{k} =& \cos \theta_{k} e ^{- i \frac{\phi_{k}}{2}} ,
\nonumber \\
\nu_{k}= & \sin \theta_{k} e ^{- i \frac{\phi_{k}}{2}} , \
\bar{\nu}_{k} =& \sin \theta_{k} e ^{ i \frac{\phi_{k}}{2}} ,
\end{eqnarray}
where $\theta _{k}$ and $\phi _{k}$ are
\begin{eqnarray*}
\phi _{k} &=& \left\{ \begin{aligned} & \frac{k}{2} + \tan ^{-1}
\left[ \delta \tan \left( \frac{k}{2} \right) \right] , & k < \pi , \\
& \frac{k}{2} + \tan ^{-1} \left[ \delta \tan \left( \frac{k}{2}
\right) \right] +\pi , & k \geq \pi , \end{aligned} \right . \\
\theta _{k} &=& \cos ^{-1} \left[ \sqrt{ \left( 1 + i \lambda _{k}
\right) /2} \right] ,
\end{eqnarray*}
with $\phi _{k}\in \left[ 0 , 2\pi \right] $, $\delta >0$ and $\lambda
_{k}=\gamma /\epsilon _{k}$.

The non-Hermitian operator $H_{k}$ in Eq.~(\ref{Dia H_k}) is in
diagonal form, since $\alpha_{k}$, $\bar{\alpha}_{k}$, $\beta _{k}$,
and $\bar{\beta}_{k}$ are canonical conjugate operators, obeying the
canonical commutation relations
\begin{eqnarray}
\left[ \alpha_{k} , \bar{\alpha}_{k^{\prime}}\right] _{\pm} = &\left[
\beta_{k} , \bar{\beta}_{k^{\prime}} \right]_{\pm} & =
\delta_{kk^{\prime}} , \nonumber \\
\left[ \alpha_{k} , \alpha_{k^{\prime}} \right]_{\pm} = & \left[
\beta_{k} , \beta_{k^{\prime}} \right]_{\pm } & = 0,  \nonumber \\
\left[ \bar{\alpha}_{k} , \bar{\alpha}_{k^{\prime}} \right] _{\pm} = &
\left[ \bar{\beta}_{k} , \bar{\beta}_{k^{\prime}} \right]_{\pm } & =
0 , \label{Canonical.CR} \\
\left[ \alpha_{k} , \bar{\beta}_{k^{\prime}} \right]_{\pm} = & \left[
\bar{\alpha}_{k} , \bar{\beta}_{k^{\prime}} \right]_{\pm } & = 0 ,
\nonumber \\
\left[ \alpha_{k} , \beta_{k^{\prime}} \right]_{\pm} = & \left[
\bar{\alpha}_{k} , \beta_{k^{\prime}} \right]_{\pm} & = 0 .  \nonumber
\end{eqnarray}
Therefore, the original Hamiltonian (\ref{Ham0}) is diagonalized. The
method employed here is similar to that for the Hermitian two-band
models \cite{PRB2099,HMXEPL,ChenB}. Nevertheless, the transformation
in Eqs.~(\ref{apha_k}) and (\ref{apha_k bar}) is no longer unitary
under the Dirac inner product, since the canonical conjugate pairs
appearing in Eq.~(\ref{Canonical.CR}) are not simply defined by the
Hermitian conjugate operation, i.e. $\bar{\alpha}_{k} \neq \alpha_{k}
^{\dagger}$ and $\bar{\beta}_{k} \neq \beta_{k}^{\dagger}$, which is
crucial in this work.

We note that the spectrum $\epsilon _{k}$ consists of two branches
separated by an energy gap
\begin{equation}
\Delta = \sqrt{4J^{2} \delta^{2} - \gamma^{2}}.
\end{equation}
Obviously, it displays a full real spectrum within the region of
$4J^{2} \delta^{2} \geq \gamma ^{2}$. Beyond this region, the
imaginary eigenvalues appears and the $\mathcal{PT}$ symmetry of the
corresponding eigenfunction is broken simultaneously according to the
non-Hermitian quantum theory. Interestingly, it occurs independently
on the size of the lattice. Notice that, when the onset of the
$\mathcal{PT}$ symmetry breaking begins, the band gap vanishes, which
is similar to that in a Hermitian two-band model. However, the
dimerization still exists ($\delta \neq 0$), when the gap vanishes in
this non-Hermitian model. In the next section, the further
relationship between a non-Hermitian and a Hermitian two band models
will be discussed.

\section{Hermitian counterpart}
\label{sec.HerCou}

In this section, we would like to construct the equivalent Hermitian
counterpart of the non-Hermitian model Eq.~(\ref{Ham0}), which is a
typical topic in the non-Hermitian quantum theory. In general, this
can be done in the framework of metric-operator theory \cite{mosta1,%
mosta2}. Nevertheless, for the present model one can achieve this goal
in a more direct way. This is due to the fact that the spectrum
$\epsilon _{k}$ has an evident physical meaning. To demonstrate this
point, we consider the model of a Peierls distorted tight-binding ring
with staggered real potentials. The Hamiltonian can be written as
\begin{eqnarray}
H_{\mathrm{e}} &=& - J_{\mathrm{e}} \sum_{l=1}^{2N} \left[ 1 + \left(
-1 \right) ^{l} \delta _{\mathrm{e}} \right] \left( b_{l}^{\dagger}
b_{l+1} + \mathrm{H.c.} \right)   \nonumber \\
&& + V_{\mathrm{e}} \sum_{l} \left( -1 \right) ^{l} b_{l}^{\dagger}
b_{l} ,
\end{eqnarray}
where $b_{l}^{\dagger}$ is the creation operator of a boson (or a
fermion) at the $l$th site, with the periodic boundary condition
$b_{2N+1}=b_{1}$. This Hamiltonian can be viewed as the Hermitian
counterpart, which will be shown in the following. By the similar
procedure, taking the unitary transformation
\begin{eqnarray}
\mathcal{A}_{k} &=& \frac{1}{\sqrt{N}} \sum_{l=1}^{N} \left( \zeta
_{k} e ^{- i kl} b_{2l-1} + \xi _{k} e ^{- i kl} b_{2l} \right) ,
\nonumber \\
\mathcal{B}_{k} &=& \frac{1}{\sqrt{N}} \sum_{l=1}^{N} \left( -\zeta
_{k}^{\ast} e ^{- i kl} b_{2l-1} + \xi _{k}^{\ast } e ^{- i kl} b_{2l}
\right) ,  \label{eq.A_k}
\end{eqnarray}
and $H_{\mathrm{e}}$ can be written in the diagonal form
\begin{equation}
H_{\mathrm{e}} = - \sum_{k} \varepsilon _{k} \left( \mathcal{A} _{k}
^{\dagger} \mathcal{A}_{k} - \mathcal{B}_{k} ^{\dagger} \mathcal{B}
_{k} \right) ,
\end{equation}
where the spectrum is
\begin{equation}
\varepsilon _{k} = 2J _{\mathrm{e}} \sqrt{ \left( 1 - \delta
_{\mathrm{e}} ^{2} \right) \cos ^{2} \left( \frac{k}{2} \right)
+ \delta _{\mathrm{e}} ^{2} + \left( \frac{V_{\mathrm{e}}} {2J
_{\mathrm{e}} } \right) ^{2}}.
\label{spectrum e}
\end{equation}
It is different from the situation of a non-Hermitian model, the
coefficients $\zeta _{k}$ and $\xi _{k}$ satisfy
\begin{equation}
\left\vert \zeta _{k} \right\vert ^{2} + \left\vert \xi _{k}
\right\vert ^{2} = 1 ,
\end{equation}
which ensures the unitarity of the transformation in
Eq.~(\ref{eq.A_k}) and the canonical commutation relation
\begin{eqnarray}
\left[ \mathcal{A}_{k} , \mathcal{A}_{k^{\prime}} ^{\dagger} \right]
_{\pm} &=& \left[ \mathcal{B}_{k} , \mathcal{B}_{k^{\prime }}
^{\dagger} \right] _{\pm} = \delta _{kk^{\prime}},  \nonumber \\
\left[ \mathcal{A}_{k} , \mathcal{B}_{k^{\prime}} ^{\dagger} \right]
_{\pm } &=& 0 .
\end{eqnarray}
Comparing two spectra $\epsilon _{k}$ and $\varepsilon _{k}$, one can
see that they can be identical under the condition
\begin{equation}
\frac{\delta ^{2} - \left( \gamma /2J \right) ^{2}}{ 1 -\delta ^{2}} =
\frac{\delta _{\mathrm{e}} ^{2} + \left( V _{\mathrm{e}} /
2J_{\mathrm{e}} \right) ^{2}}{1 - \delta _{\mathrm{e}}^{2}}.
\end{equation}
Therefore, Hamiltonian $H_{\mathrm{e}}$ can be regarded as an
equivalent Hermitian Hamiltonian of $H$.

To illustrate this point, we consider a simple case of
$H_{\mathrm{e}}$ with no energy gap $\Delta =0$ and $\gamma =
\gamma_{\mathrm{c}} = 2J\delta$. Then the corresponding equivalent
Hermitian Hamiltonian has the form
\begin{equation}
h_{\mathrm{e}} = -J_{\mathrm{e}} \sum_{j=1}^{2N} \left( b_{j}
^{\dagger} b_{j+1} + \mathrm{H.c.} \right) .
\label{eq.H_eff}
\end{equation}
which represents a uniform ring system with hopping amplitude
$J_{\mathrm{e}} = J \sqrt{1 - \delta^{2}}$. In Sec.~\ref{sec.quasi} we
will investigate the wave-packet dynamics. It is noted that, although
the spectrum for the non-Hermitian model is equivalent to that of a
uniform ring, the distortions $\delta$ and the imaginary potentials
$\gamma$ are still nonzero and affect the dynamics in a balanced
manner.

We would like to point out that the method employed in this work is
not universal as it depends on the obtained spectrum. We believe that
the equivalent Hamiltonian $H_{\mathrm{e}}$ can be obtained by the
standard metric-operator theory \cite{mosta1,mosta2}. Actually, both
methods have been used to another non-Hermitian model in a previous
work \cite{ZXZ}.

\section{Quasi orthogonality and Hermitian dynamics}
\label{sec.quasi}

It is well known that the eigenstates of a non-Hermitian Hamiltonian
can construct a set of biorthogonal bases in associate with the
eigenstates of its Hermitian conjugate. For the present Hamiltonian in
Eq.~(\ref{Ham0}), eigenstates \{$\bar{\alpha}_{k}\left\vert 0
\right\rangle , \bar{\beta}_{k} \left\vert 0 \right\rangle $\} of $H$
and eigenstates \{$\alpha_{k}^{\dagger }\left\vert 0\right\rangle ,
\beta_{k} ^{\dagger} \left\vert 0 \right\rangle $\} of $H^{\dagger}$
are the biorthogonal bases of the single-particle invariant subspace.
This can be extended to the many-particle invariant subspace due to
the canonical commutation relations in Eq.~(\ref{Canonical.CR}). On
the other hand, the eigenstates of a non-Hermitian Hamiltonian are not
orthogonal under the Dirac inner product in the general case. However,
we note that the eigenstates of the present Hamiltonian (\ref{Ham0})
are the eigenstates of momentum simultaneously, which should lead to
the orthogonality between the eigenstates with different $k$ in the
Dirac inner product. This property is reflected by the following
quasi-canonical commutation relations
\begin{eqnarray}
\left[ \alpha_{k} , \alpha_{k^{\prime}}^{\dagger} \right]_{\pm} =
& \left[ \beta_{k} , \beta_{k^{\prime}}^{\dagger} \right]_{\pm} & =
\sqrt{1 + \lambda_{k}^{2}} \delta_{kk^{\prime}},  \nonumber \\
\left[ \bar{\alpha}_{k}^{\dagger} , \bar{\alpha}_{k^{\prime}} \right]
_{\pm} = & \left[ \bar{\beta}_{k}^{\dagger} , \bar{\beta}_{k^{\prime}}
\right] _{\pm } & = \sqrt{1+\lambda _{k}^{2}}\delta _{kk^{\prime }} ,
\nonumber \\
\left[ \beta_{k} , \alpha_{k^{\prime}}^{\dagger} \right]_{\pm} =
& \left[ \bar{\alpha}_{k}^{\dagger} , \bar{\beta}_{k^{\prime}} \right]
_{\pm} & = i \lambda_{k} \delta_{kk^{\prime}},  \label{quasi.CR} \\
\left[ \alpha_{k} , \bar{\alpha}_{k^{\prime}}^{\dagger} \right]_{\pm}
=& \left[ \beta_{k} , \bar{\beta}_{k^{\prime}}^{\dagger} \right]_{\pm}
& =0, \nonumber \\
\left[ \alpha_{k} , \bar{\beta}_{k^{\prime}}^{\dagger} \right]_{\pm }
=& \left[ \beta_{k} , \bar{\alpha}_{k^{\prime}}^{\dagger} \right]
_{\pm} & =0 . \nonumber
\end{eqnarray}
Here the term ``quasi'' is the manifestation of the non-Hermitian
nature of $H$ in Eq.~(\ref{Canonical.CR}), which is represented in the
absence of orthogonality between the eigenmodes of $\bar{\alpha}_{k}$
and $\bar{\beta}_{k}$. On the other hand, the rest ``canonical
commutation relations'' makes the non-Hermitian system appear
Hermitian to some extent. Similar relations and corresponding
dynamical phenomena in a $\mathcal{PT}$-symmetric ladder system were
presented in a previous work \cite{LJin84}.

Now we turn to investigate the dynamics of such two-band model. Owing
to the non-Hermiticity of the Hamiltonian, the time evolution operator
$U \left( t \right) =\exp \left( - i Ht \right)$ is not unitary. To
clarify the feature of the dynamics, we consider the time evolution of
an arbitrary state. For the given initial state
\begin{equation}
\left\vert \psi \left( 0\right) \right\rangle = \sum_{k} \left( f_{k}
\bar{\alpha}_{k} + g_{k} \bar{\beta}_{k} \right) \left\vert 0
\right\rangle ,
\end{equation}
we have
\begin{eqnarray}
\left\vert \psi \left( t\right) \right\rangle &=& U\left( t\right)
\left\vert \psi \left( 0\right) \right\rangle  \nonumber \\
&=& \sum_{k}\left( e ^{ i \epsilon_{k}t} f_{k} \bar{\alpha}_{k} + e ^{
- i \epsilon_{k}t} g_{k} \bar{\beta}_{k} \right) \left\vert 0
\right\rangle .
\end{eqnarray}
There are two types of probability, $P_{\mathrm{D}}\left( t\right)$
and $P_{\mathrm{B}}\left( t\right)$, in terms of the Dirac and
biorthogonal inner product, respectively, i.e.
\begin{eqnarray}
P_{\mathrm{D}}\left( t\right) &=& \left\vert U \left( t\right)
\left\vert \psi \left( 0\right) \right\rangle \right\vert
_{\mathrm{D}}^{2} \\
&=& \sum_{k}\left\langle 0\right\vert \left( e ^{- i \epsilon _{k} t}
f_{k}^{\ast} \bar{\alpha}_{k}^{\dagger} + e ^{ i \epsilon_{k}t}
g_{k}^{\ast } \bar{\beta}_{k}^{\dagger} \right) U \left( t\right)
\left\vert \psi \left( 0\right) \right\rangle ,  \nonumber
\end{eqnarray}
\begin{eqnarray}
P_{\mathrm{B}}\left( t\right) &=& \left\vert U\left( t\right)
\left\vert \psi \left( 0\right) \right\rangle \right\vert
_{\mathrm{B}}^{2} \\
&=& \sum_{k}\left\langle 0\right\vert \left( e ^{- i \epsilon_{k}t}
f_{k}^{\ast} \alpha_{k} + e ^{ i \epsilon_{k}t} g_{k}^{\ast} \beta_{k}
\right) U\left( t\right) \left\vert \psi \left( 0\right) \right\rangle
, \nonumber
\end{eqnarray}
where $\left\vert \left\vert \psi \right\rangle \right\vert
_{\mathrm{D}} ^{2}$ and $\left\vert \left\vert \psi \right\rangle
\right\vert _{\mathrm{B}}^{2}$ denote the Dirac and biorthogonal norms
of the state $\left\vert \psi \right\rangle$, respectively. From the
commutation relations Eq.~(\ref{Canonical.CR}), we have
$P_{\mathrm{B}}\left( t\right) =1$, which is the aim of the
introduction of the biorthogonal inner product. In contrast,
$P_{\mathrm{D}}\left( t\right)$ is not unity and probably huge in some
cases \cite{ZXZ}.

From the quasi-canonical commutation relations of
Eq.~(\ref{quasi.CR}), we have
\begin{eqnarray}
P_{\mathrm{D}}\left( t\right) &=& \sum_{k}\left( \left\vert f_{k}
\right\vert^{2} + \left\vert g_{k} \right\vert^{2} \right) \sqrt{ 1 +
\lambda_{k}^{2}}  \nonumber \\
&& + 2 \sum_{k} \lambda_{k} \left\vert g_{k} f_{k} \right\vert \sin
\left( 2 \epsilon_{k} t + \varphi_{k}\right) ,  \label{P_D}
\end{eqnarray}
where $\lambda_{k} = \gamma / \epsilon_{k}$ and $\varphi_{k}$ is a
time-independent phase defined as $ e ^{ i \varphi_{k}} = g_{k}^{\ast}
f_{k} / \left\vert g_{k} f_{k} \right\vert$. Obviously, the first term
is time-independent while the second term represents a summation of
periodic sinusoidal functions with frequency $2\epsilon _{k}$. In case
of $g_{k} f_{k} =0$ (for each eigenmode $k$, the initial state does
not comprise components of $\bar{\alpha}_{k}$ and $\bar{\beta}_{k}$
simultaneously) and $\lambda_{k}$ being finite (the initial state does
not comprise the component of $\epsilon_{k}=0$, when the Hamiltonian
becomes a Jordan block operator), the probability-preserving time
evolution occurs. Nevertheless, even in the case of $g_{k} f_{k} \neq
0$, if $\lambda_{k} \ll 1$, the probability slightly fluctuates around
a certain constant, with the time evolution being
quasi-probability-preserving.

\section{Wave packet dynamics}
\label{sec.EffRing}

Now we apply the obtained results to a more concrete case and then
demonstrate the dynamic property of the system. We investigate the
time evolution of the wave packet in the system with zero band gap. As
mentioned above, it has been shown that the spectrum of the system is
the same as that of a uniform ring, which can be regarded as the
equivalent Hermitian counterpart.

\begin{figure*}[tbph]
\includegraphics[bb=0 0 567 684,width=0.325\textwidth,clip]{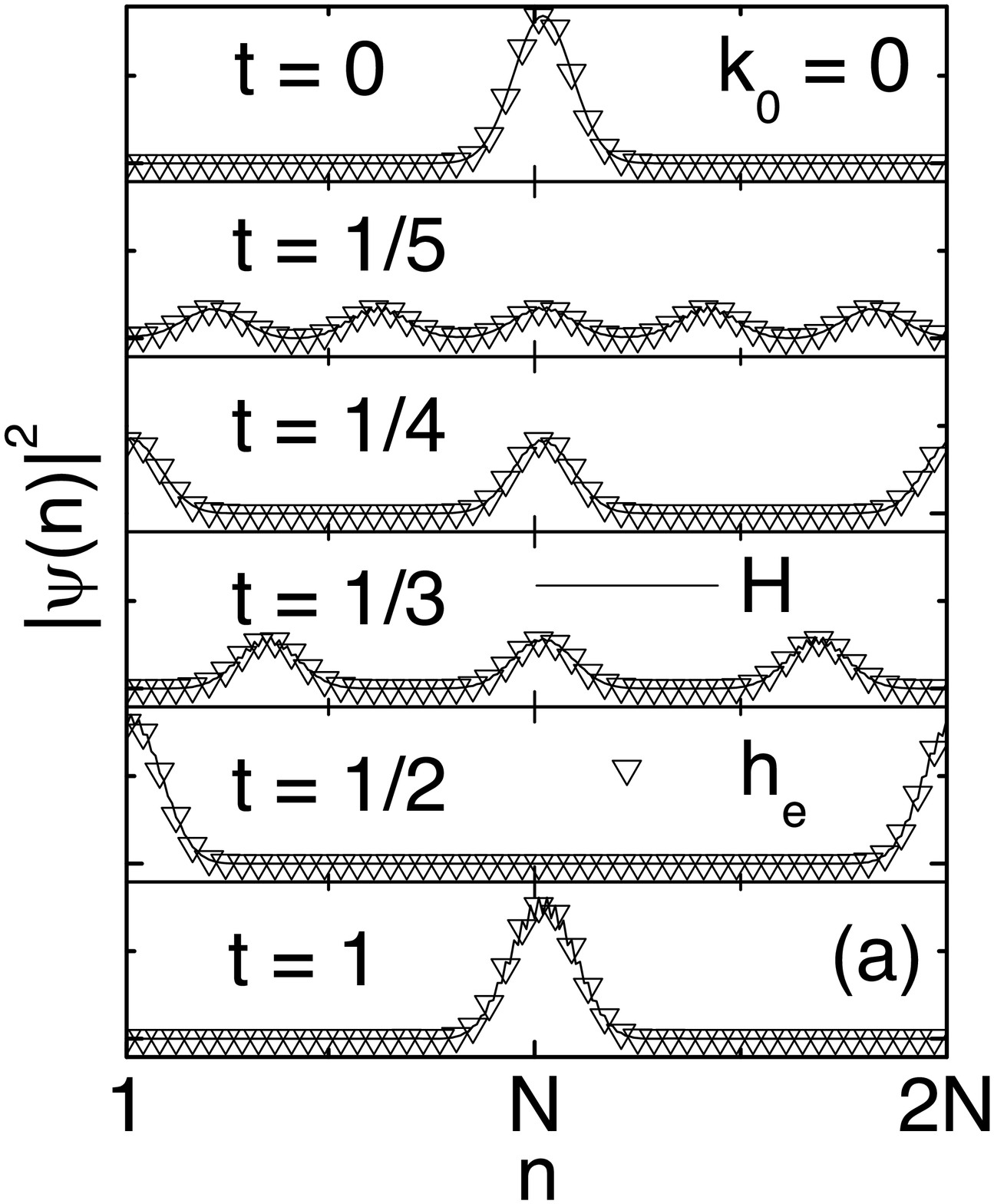}
\includegraphics[bb=0 0 567 683,width=0.325\textwidth,clip]{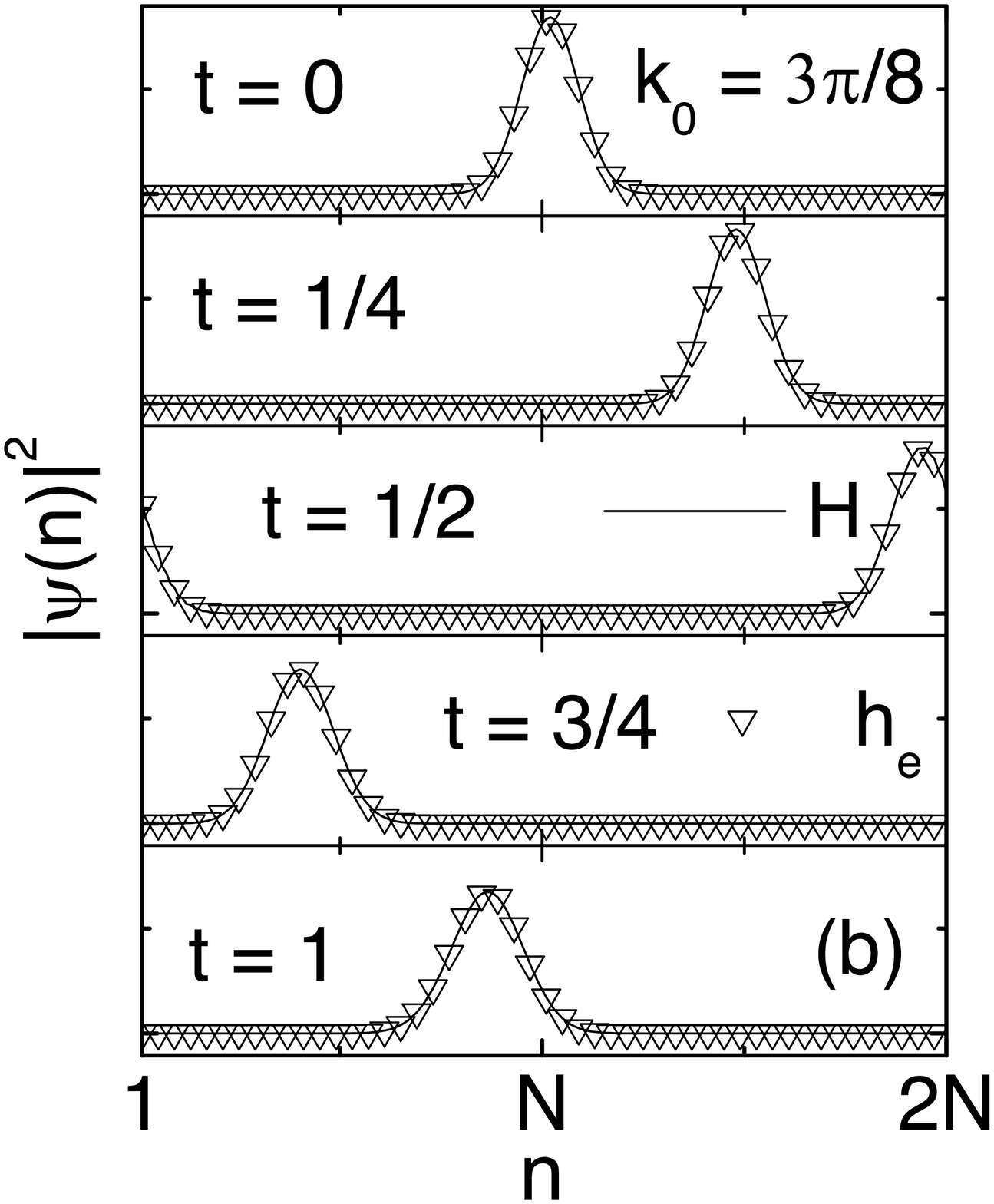}
\includegraphics[bb=0 0 567 683,width=0.325\textwidth,clip]{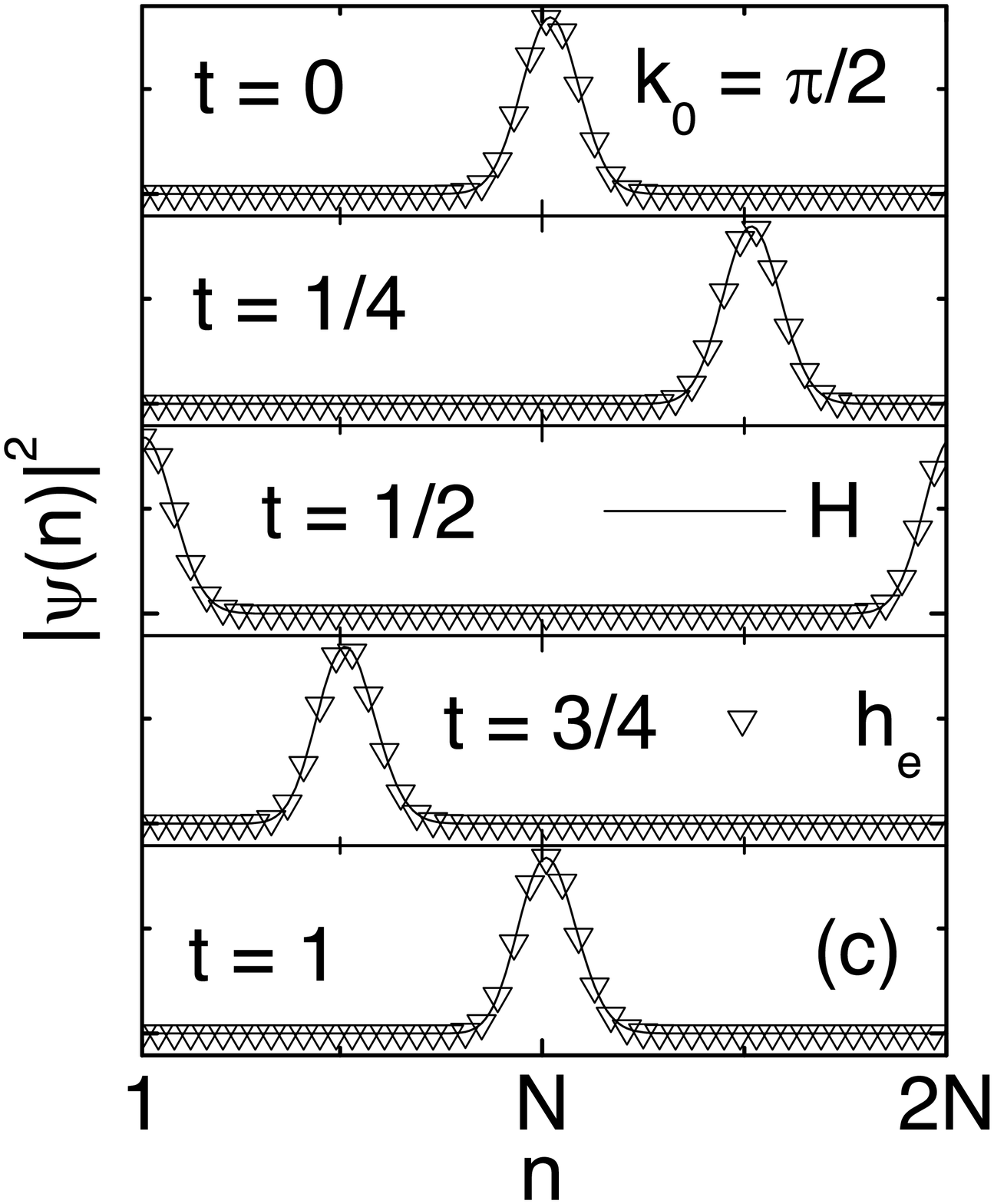}
\caption{The illustration of the time evolution of a Gaussian wave
packet (solid line) with $\alpha =0.1$ and (a) $k_{0}=0$ (b) $k_{0} =
3\pi /8$ (c) $k_{0}=\pi /2$ in a ring of $N=100$, $\delta =0.1$ and
$\gamma = 0.2 - 10^{-8} \sim \gamma_{\mathrm{c}}$ (in units of $J$),
where $\gamma_{\mathrm{c}} = 2J \delta$. We take $t$ in units of
$T_{\mathrm{rev}}$ from Eq.~(\ref{eq.T_rev}) in (a), and
$T_{\mathrm{cir}}$ from Eq.~(\ref{eq.T_cir}) in (b) and (c). For
comparison we also plot the same wave packet (hallow triangle), which
evolves in a uniform ring of $h_{\mathrm{e}}$ from
Eq.~(\ref{eq.H_eff}). One can see that the wave packet of $k_{0}=0$
splits into several sub-GWPs, which almost have the same shape as the
initial one and are referred as the fractional revivals. And those of
$k_{0}=3\pi /8$ and $\pi /2$ translate smoothly in the ring, where the
latter behaves the non-spreading propagation. These figures show that
the time evolution of the GWPs under the non-Hermitian $H$ gives the
quasi-Hermitian dynamical behaviors, which is similar to that under
$h_{\mathrm{e}}$.}
\label{fig.HC_Ex}
\end{figure*}
\begin{figure*}[tbph]
\includegraphics[bb=0 0 568 398,width=0.325\textwidth,clip]{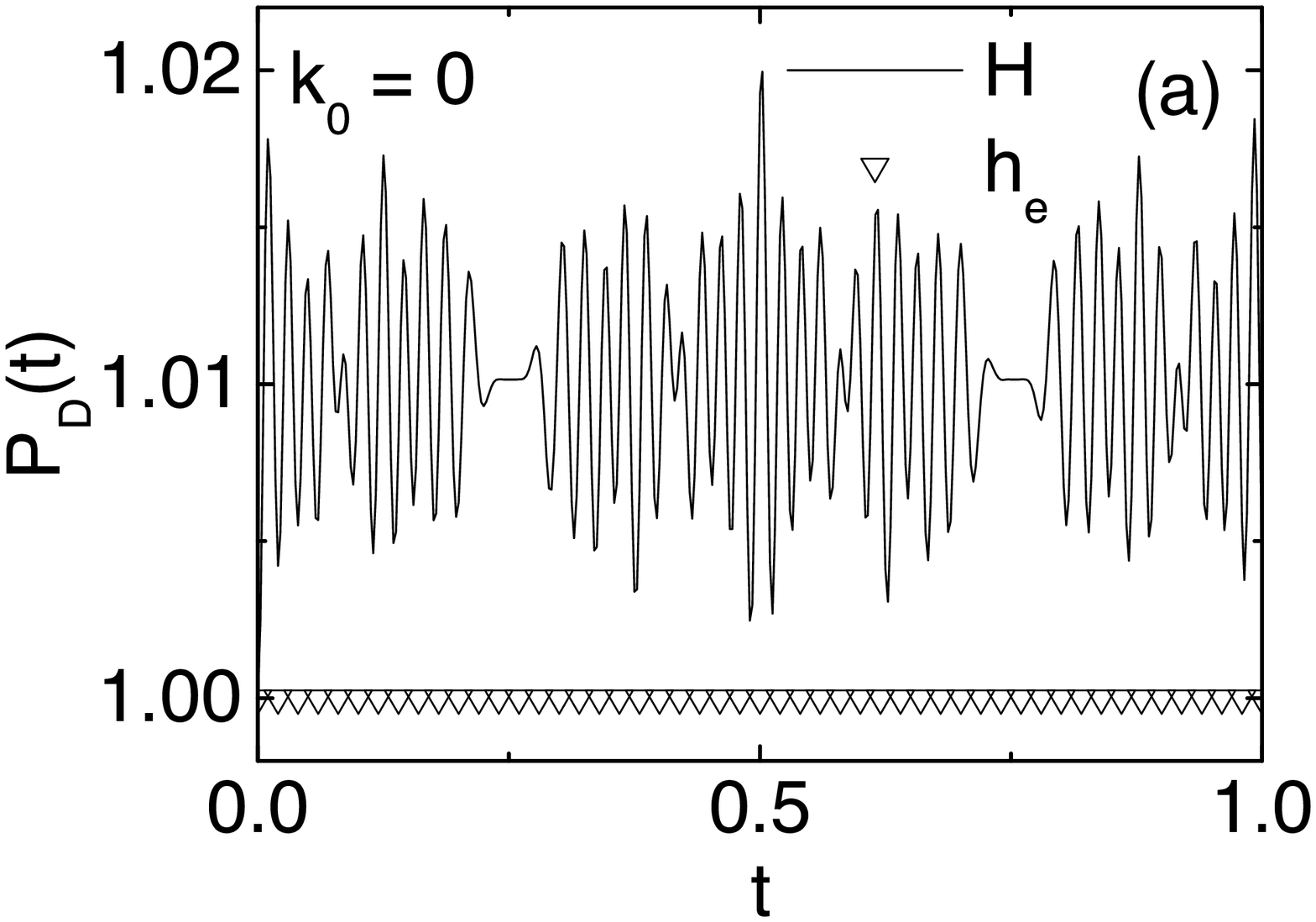}
\includegraphics[bb=0 0 568 398,width=0.325\textwidth,clip]{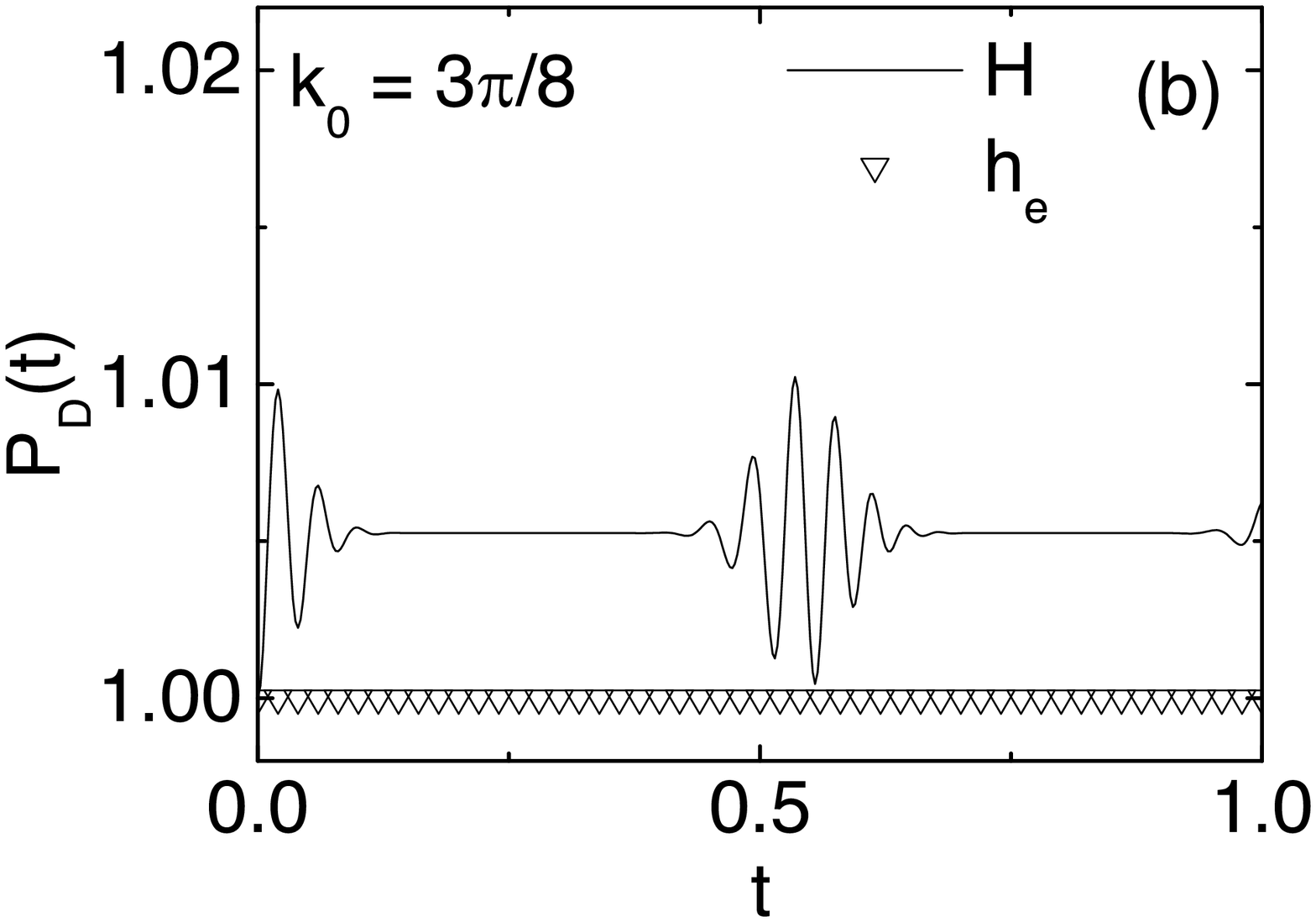}
\includegraphics[bb=0 0 568 398,width=0.325\textwidth,clip]{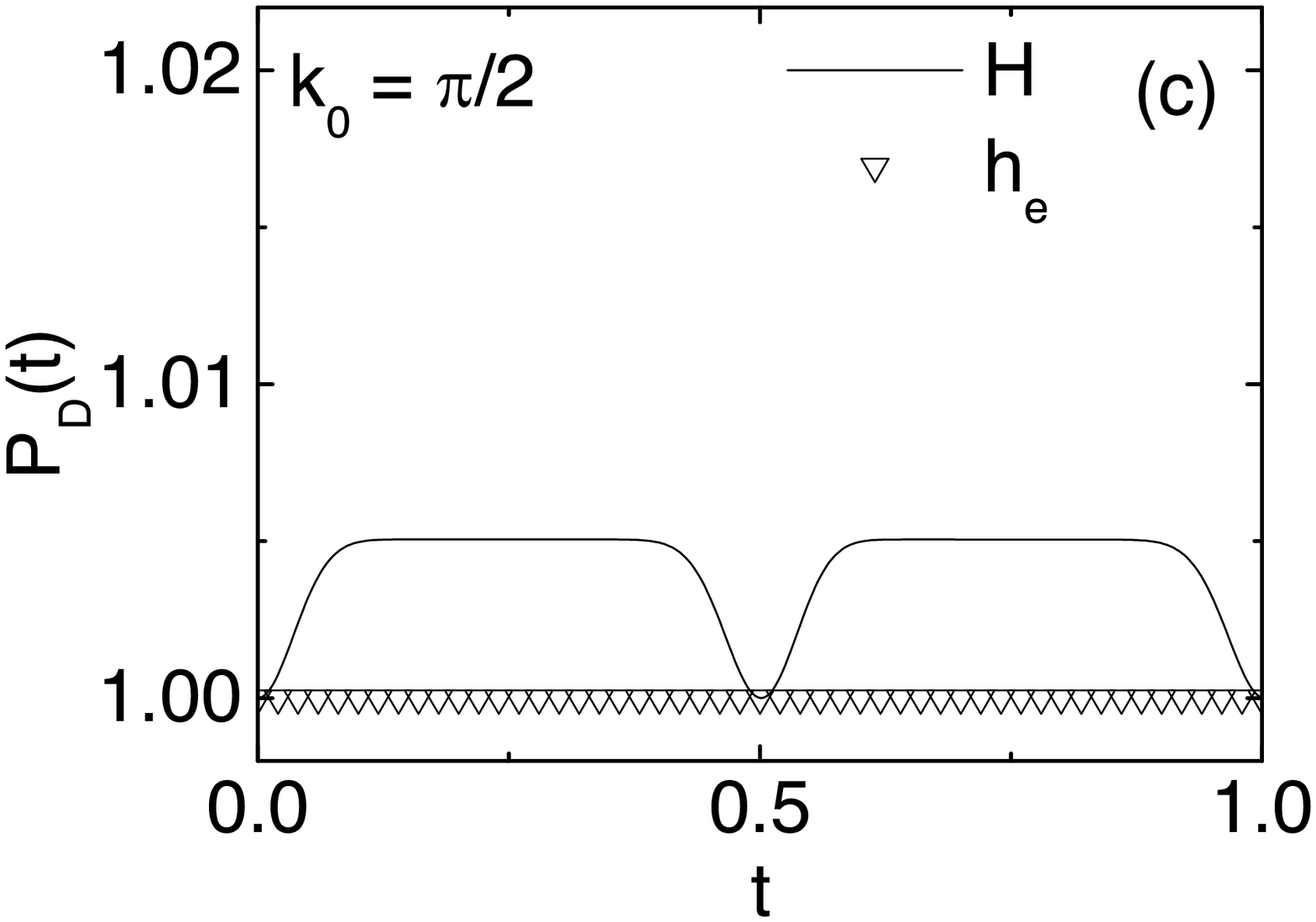}
\caption{Plots of the square of the Dirac norm (solid line) from
Eq.~(\ref{P_D}) in Fig.~\ref{fig.HC_Ex}. For comparison we also plot
the same wave packet (hallow triangle) in a uniform ring of
$h_{\mathrm{e}}$ with the hopping amplitude of $J_{\mathrm{e}}$. We
take $t$ in units of $T_{\mathrm{rev}}$ from Eq.~(\ref{eq.T_rev}) in
(a), and $T_{\mathrm{cir}}$ from Eq.~(\ref{eq.T_cir}) in (b) and (c).
One can see that the Dirac norm fluctuates slightly and deviates
little from unity. This is an obvious quasi-Hermitian dynamical
behavior, which is in agreement with our predictions.}
\label{fig.HP_Ex}
\end{figure*}

\begin{figure}[tbph]
\includegraphics[bb=0 0 568 457,width=0.35\textwidth,clip]%
{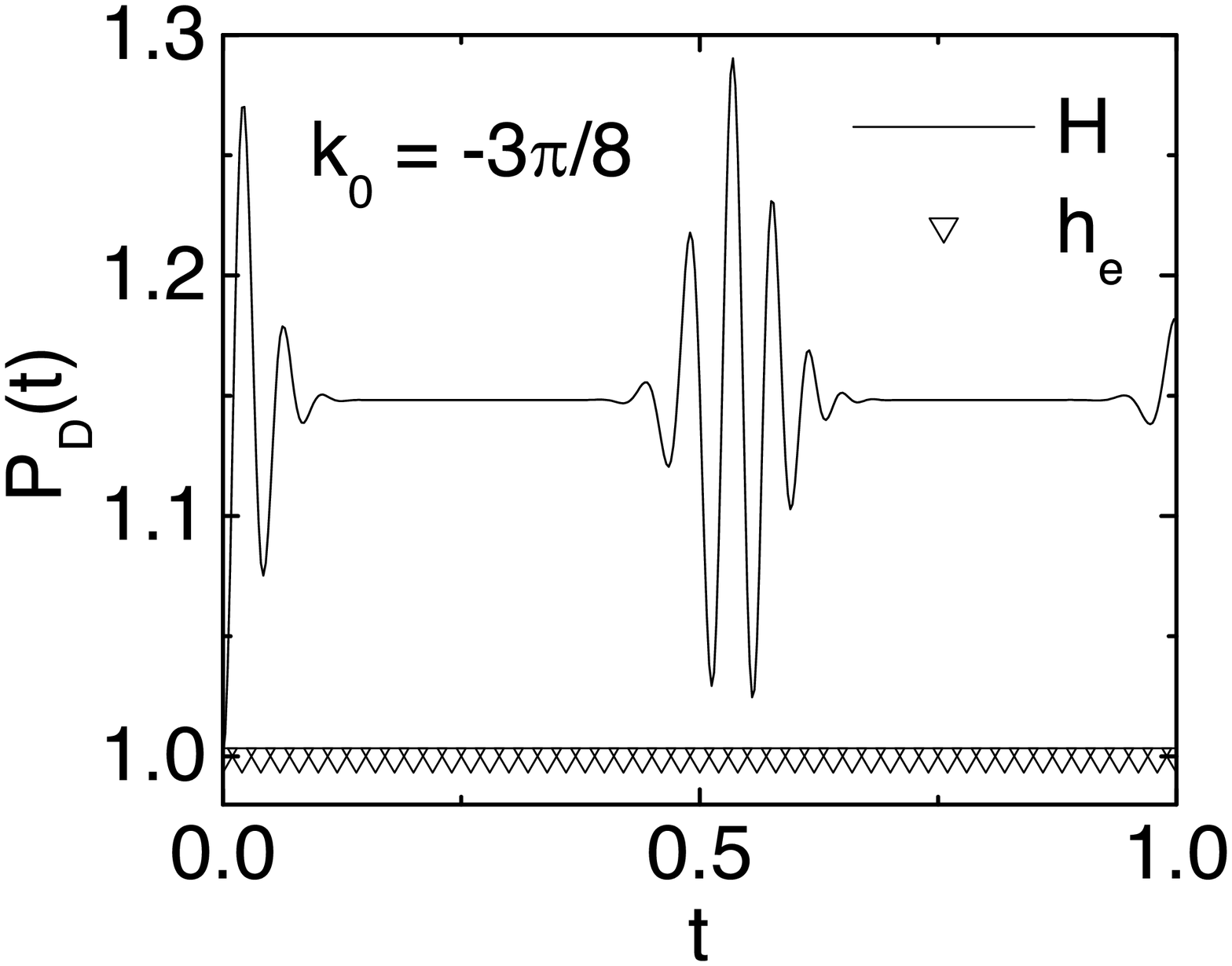}
\includegraphics[bb=0 0 568 465,width=0.35\textwidth,clip]%
{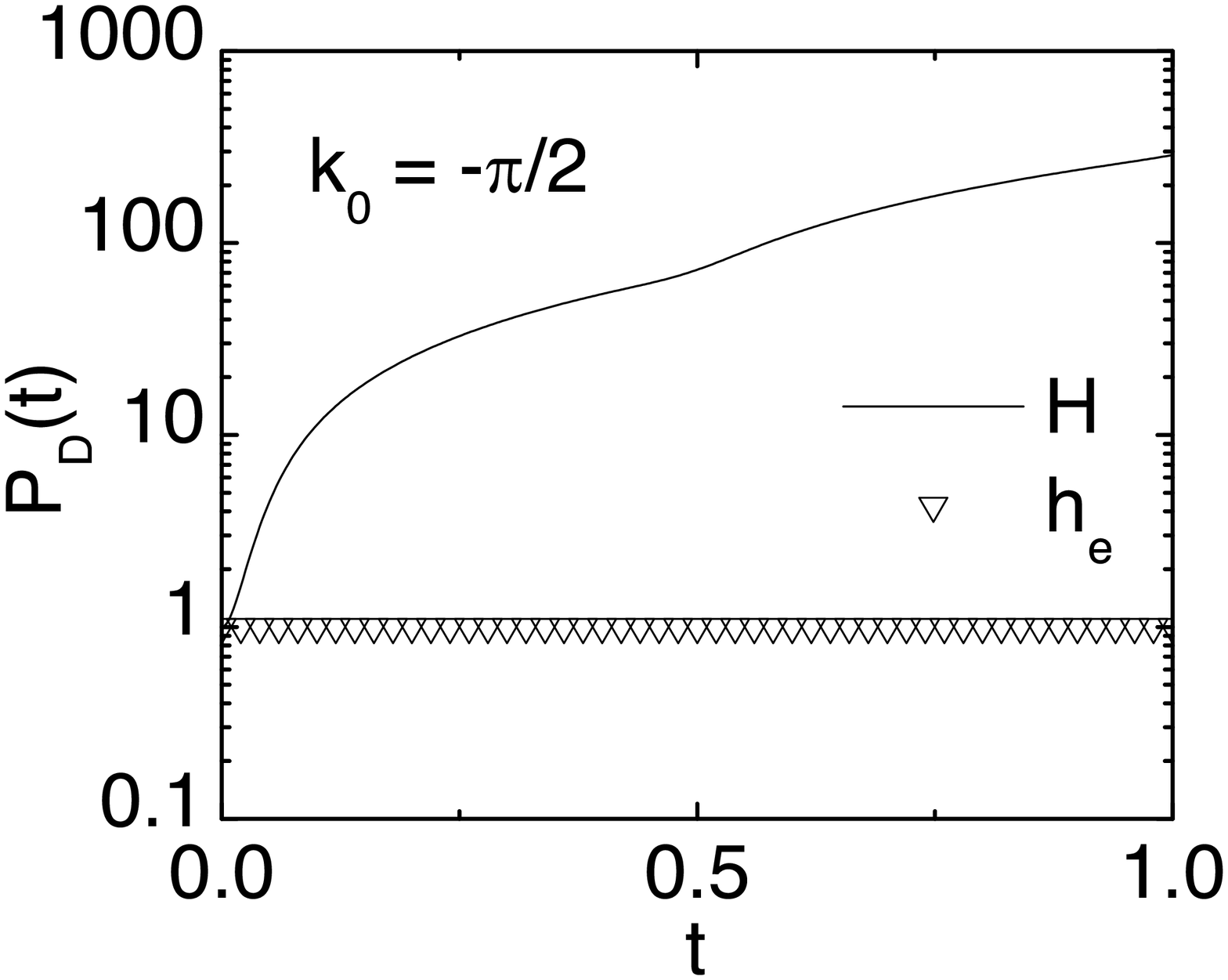}
\caption{Plots of the square of the Dirac norm (solid line) of the
evolved Gaussian wave packet with $\alpha = 0.1$, $k_{0}=-3\pi/8$
and $-\pi /2$ in the same ring as Fig.~\ref{fig.HC_Ex}. We take $t$ in
units of $T_{\mathrm{cir}}$ from Eq.~(\ref{eq.T_cir}). For comparison
we also plot the same wave packet (hallow triangle) as above in the
uniform ring of $h_{\mathrm{e}}$. One can see that the square of the
Dirac norm fluctuates greatly and its average deviates from unity
evidently, displaying evident non-Hermitian behavior, which is in
agreement with our predictions.}
\label{fig.NHP_Ex}
\end{figure}

\begin{figure}[tbph]
\includegraphics[bb=0 0 568 450,width=0.35\textwidth,clip]{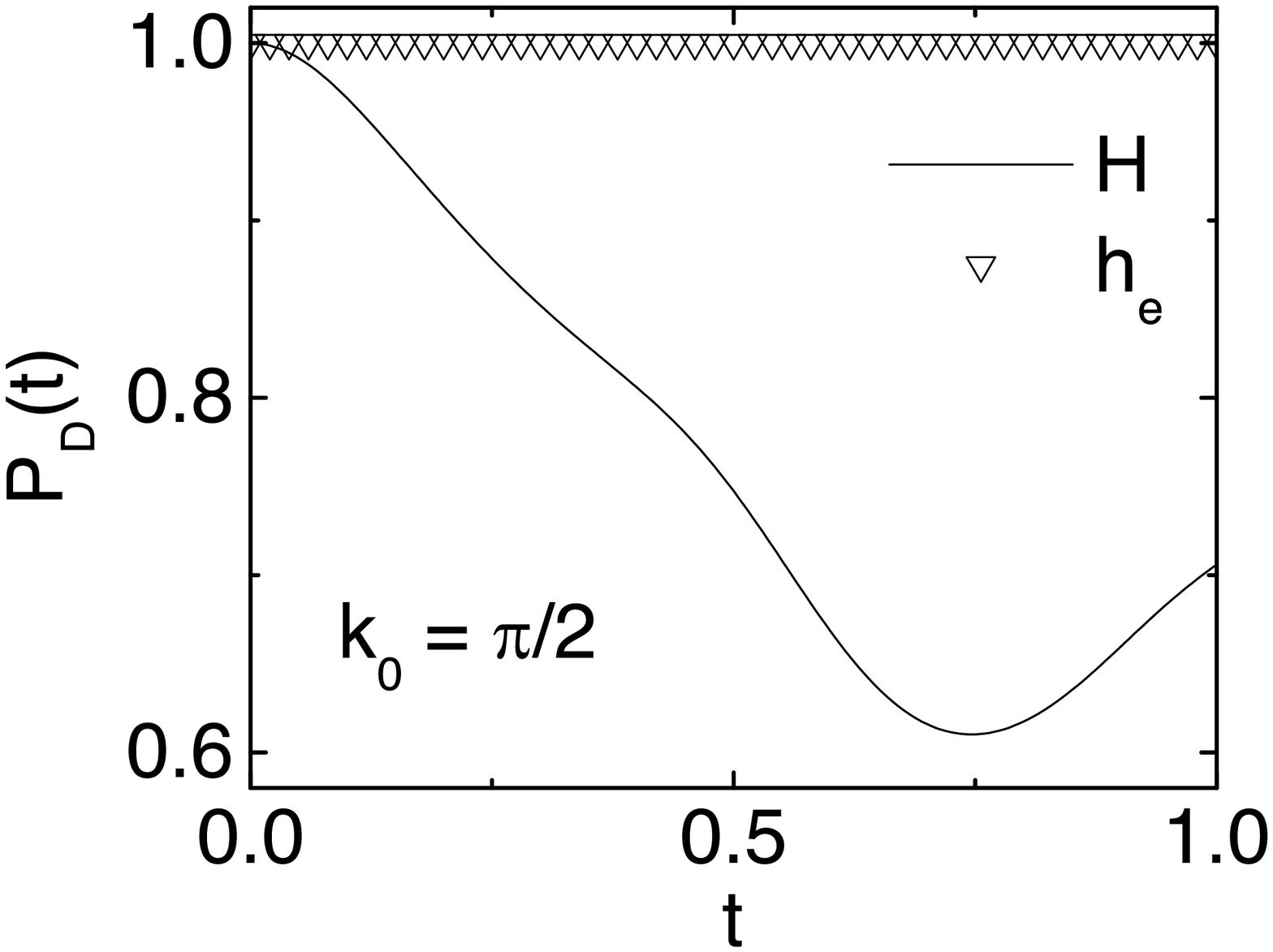}
\includegraphics[bb=0 0 568 465,width=0.35\textwidth,clip]%
{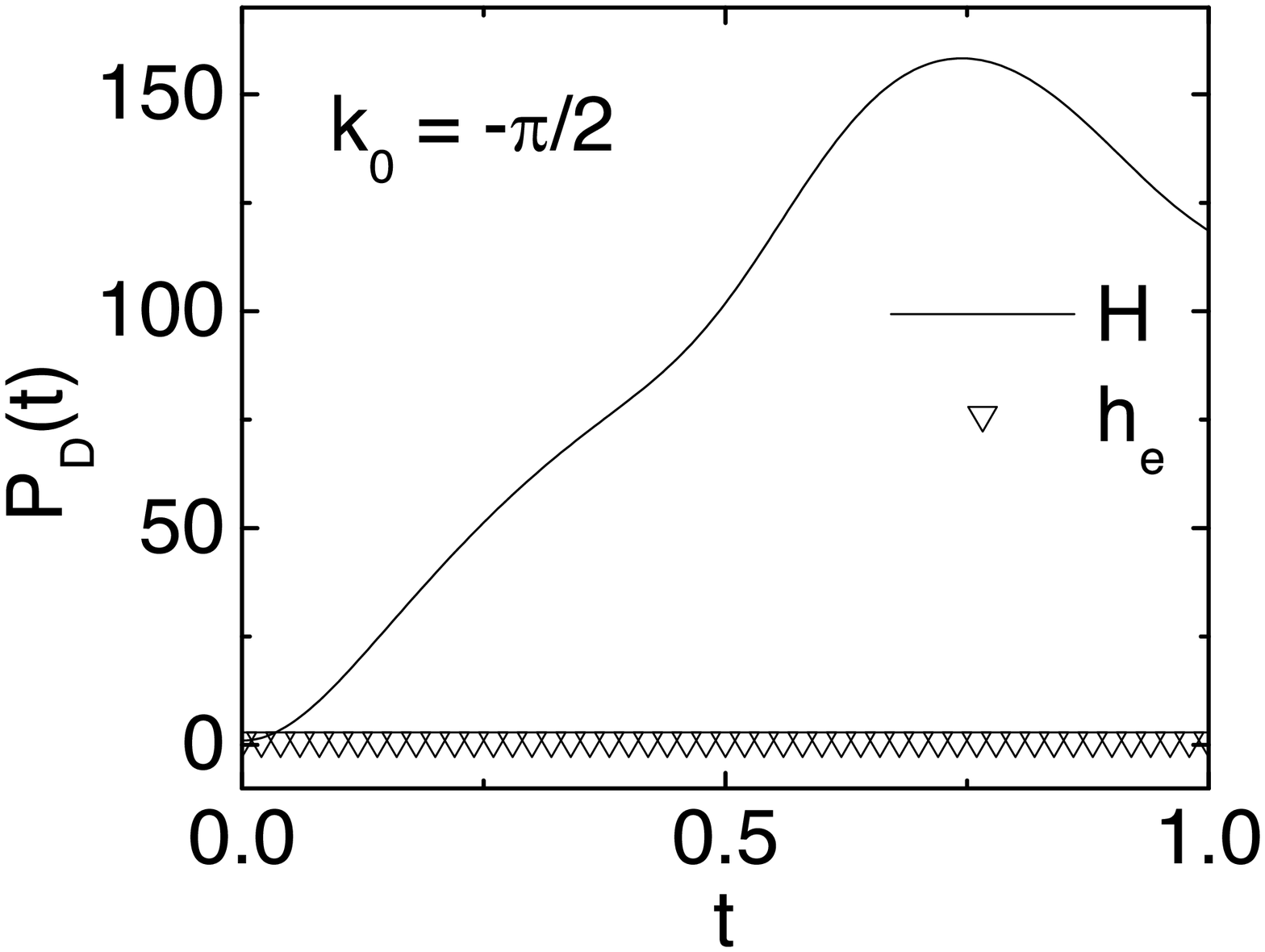}
\caption{Plots of the square of the Dirac norm (solid line) of the
evolved Gaussian wave packet with $\alpha = 0.1$, $k_{0}=\pi/2$ and
$-\pi/2$ in the same ring as Fig.~\ref{fig.HC_Ex}, but with $\gamma =
0.19$ (in units of $J$). We take $t$ in units of $T_{\mathrm{cir}}$
from Eq.~(\ref{eq.T_cir}). For comparison we also plot the same wave
packet (hallow triangle) as above in a uniform ring of
$h_{\mathrm{e}}$. One can see that the square of the Dirac norm
fluctuates greatly and its average deviates from unity evidently,
displaying obvious non-Hermitian behavior.}
\label{fig.NHP_NonEx}
\end{figure}

As an application of the obtained result, we consider the time
evolution of a Gaussian wave packet (GWP)
\begin{equation}
\left\vert \Phi \left(k_{0} , N_{\mathrm{A}} , 0 \right) \right\rangle
= \frac{1} {\sqrt{\Omega_{1}}} \sum_{l=1}^{2N} e
^{-\frac{^{\alpha^{2}}}{2} (l-N_{\mathrm{A}})^{2}} e ^{ i k_{0}l}
\left\vert l \right\rangle  \label{gaussian}
\end{equation}
with the central momentum $k_{0}\in \left[ -\pi ,\pi \right] $,
centered at the $N_{\mathrm{A}}$th site, where $\left\vert l
\right\rangle = a_{l}^{\dagger} \left\vert 0\right\rangle $ and
$\Omega _{1}$ is the normalization factor. By using the inverse
transformation from the combination of Eqs.~(\ref{A_k}) and
(\ref{apha_k})
\begin{eqnarray*}
a_{2l-1} ^{\dagger} &=& \frac{1}{\sqrt{N}} \sum_{k} e ^{ - i kl}
\left( \mu_{k} \bar{\alpha}_{k} - \bar{\nu}_{k} \bar{\beta}_{k}
\right) , \\
a_{2l} ^{\dagger} &=& \frac{1}{\sqrt{N}} \sum_{k} e ^{ - i kl} \left(
\nu_{k} \bar{\alpha}_{k} + \bar{\mu}_{k} \bar{\beta}_{k} \right) ,
\end{eqnarray*}
the GWP of Eq.~(\ref{gaussian}) has the form
\begin{eqnarray}
\left\vert \Phi (k_{0},N_{\mathrm{A}},0)\right\rangle &=& \Lambda
\sum_{k} e ^{-\frac{1}{8\alpha ^{2}}\left( k-2k_{0}\right) ^{2}} e ^{
- i N_{\mathrm{A}} \frac{k}{2}}  \nonumber \\
&& \times \left[ \eta_{k}^{+} \bar{\alpha}_{k} + \eta_{k}^{-}
\bar{\beta} _{k} \right] \left\vert 0\right\rangle ,
\end{eqnarray}
where $\Lambda = e ^{ i N_{\mathrm{A}}k_{0}} \sqrt{\pi /\left( 4
\alpha^{2} N \Omega_{1} \right)}$ and
\begin{equation}
\eta_{k}^{\pm} = \pm e ^{ i \frac{\phi_{k}}{2}} e ^{- i \frac{k}{2}}
\sqrt{1 \pm i \lambda_{k}} + e ^{- i \frac{\phi_{k}}{2}}\sqrt{1 \mp i
\lambda_{k}} ,
\end{equation}
with $\eta_{k}^{-} = -\left( \eta_{2 \pi -k}^{+} \right) ^{\ast}$. It
is a coherent superposition of eigenstates around $k\sim 2k_{0}$ in
each band. However, in the case of $\left\vert k_{0}+\pi /2\right\vert
\gg 0$, we have
\begin{eqnarray}
\left\vert \Phi \left( k_{0},N_{\mathrm{A}},0\right) \right\rangle &
\approx & \Lambda \sum_{k} e ^{-\frac{1}{8\alpha ^{2}} \left( k -
2k_{0} \right) ^{2}} e ^{- i N_{\mathrm{A}} \frac{k}{2}} \\
&& \times \left\{
\begin{aligned}
& \eta _{k}^{+} \bar{\alpha} _{k} \left\vert 0 \right\rangle , \ -
\frac{\pi}{2} < k_{0} < \frac{\pi}{2} \\
& \eta _{k}^{-} \bar{\beta} _{k} \left\vert 0 \right\rangle , \  k_{0}
< - \frac{\pi }{2} \  \mathrm{or} \  k_{0} > \frac{\pi }{2}
\end{aligned}
\right . \nonumber
\end{eqnarray}
Obviously, it satisfies the above mentioned probability-preserving
condition of $g_{k}f_{k}=0$, and then evolves as if in a uniform ring.
On the contrary, in the case of $k_{0}\sim -\pi /2$, we have
$\left\vert \eta_{k}^{+} / \eta_{k}^{-} \right\vert \approx 1$, i.e.
two eigenmodes $\bar{\alpha}_{k}$ and $\bar{\beta}_{k}$ are both the
main components of the state simultaneously. From Eq.~(\ref{P_D}) it
is predicted that the dynamics of the wave packet should show
extremely non-Hermitian behaviors. To demonstrate and confirm the
analysis, we consider two typical cases of $k_{0}=0$ and $\pi /2$.

\begin{figure*}[tbph]
\includegraphics[bb=0 0 567 697,width=0.325\textwidth,clip]{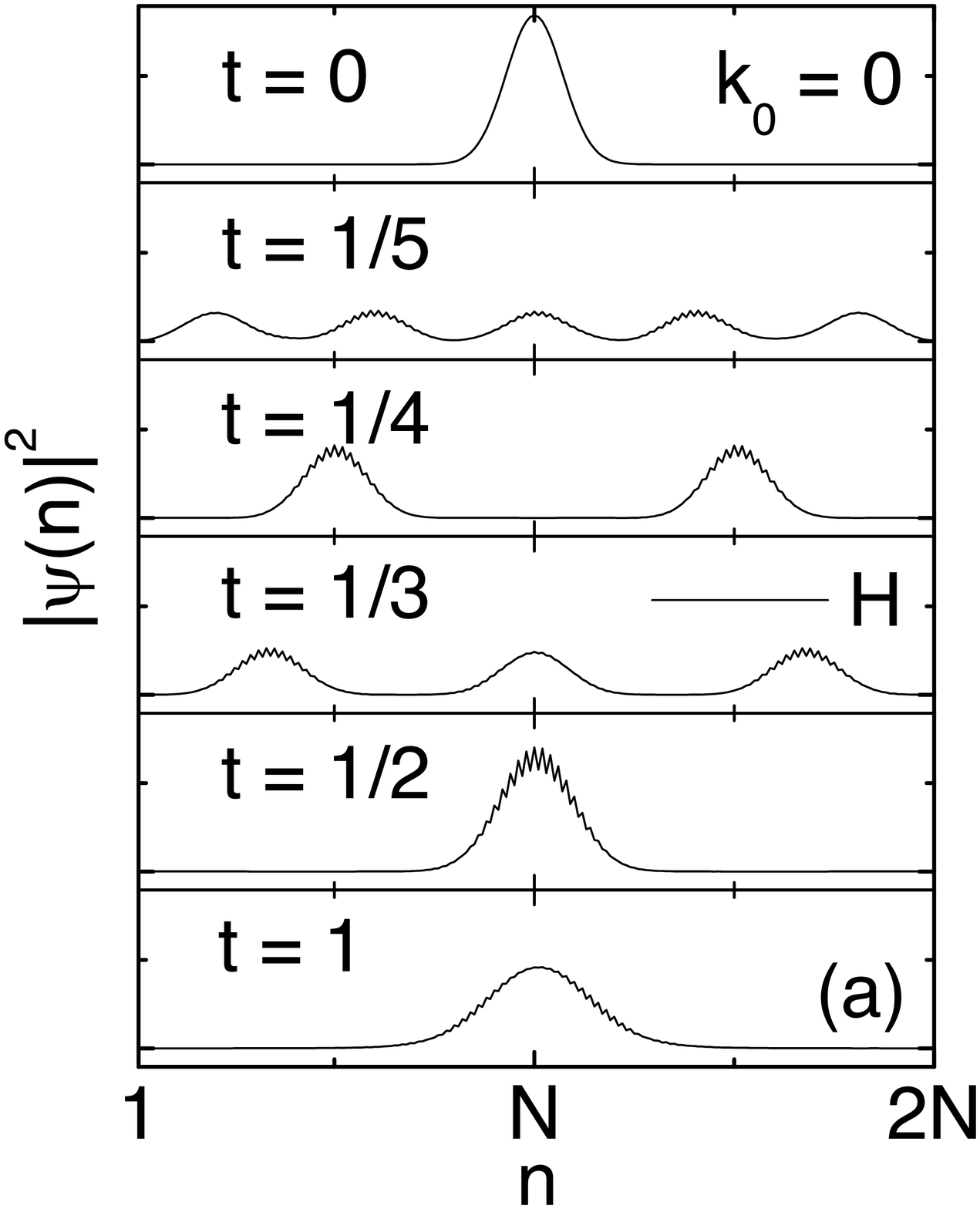}
\includegraphics[bb=0 0 567 699,width=0.325\textwidth,clip]%
{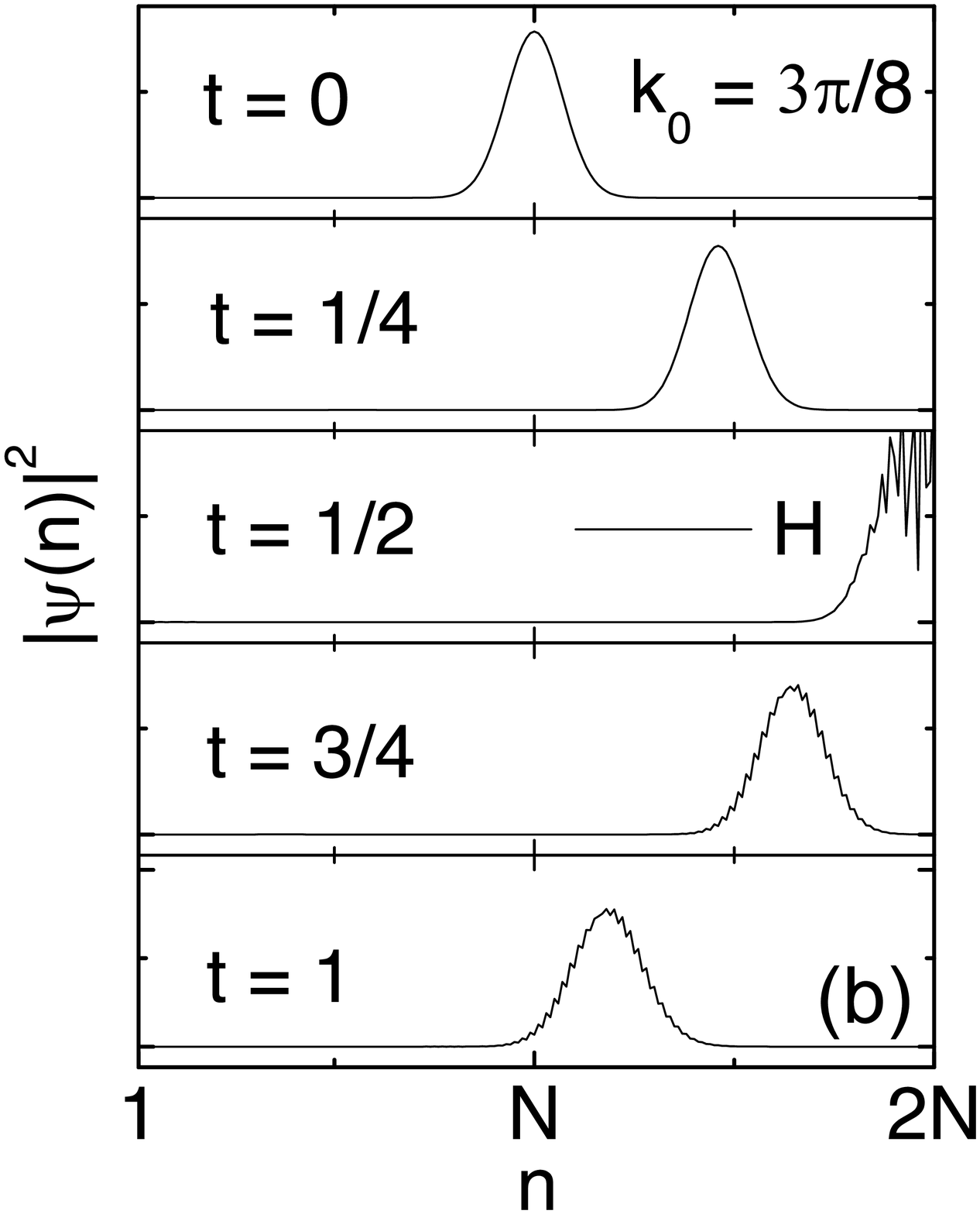}
\includegraphics[bb=0 0 567 699,width=0.325\textwidth,clip]{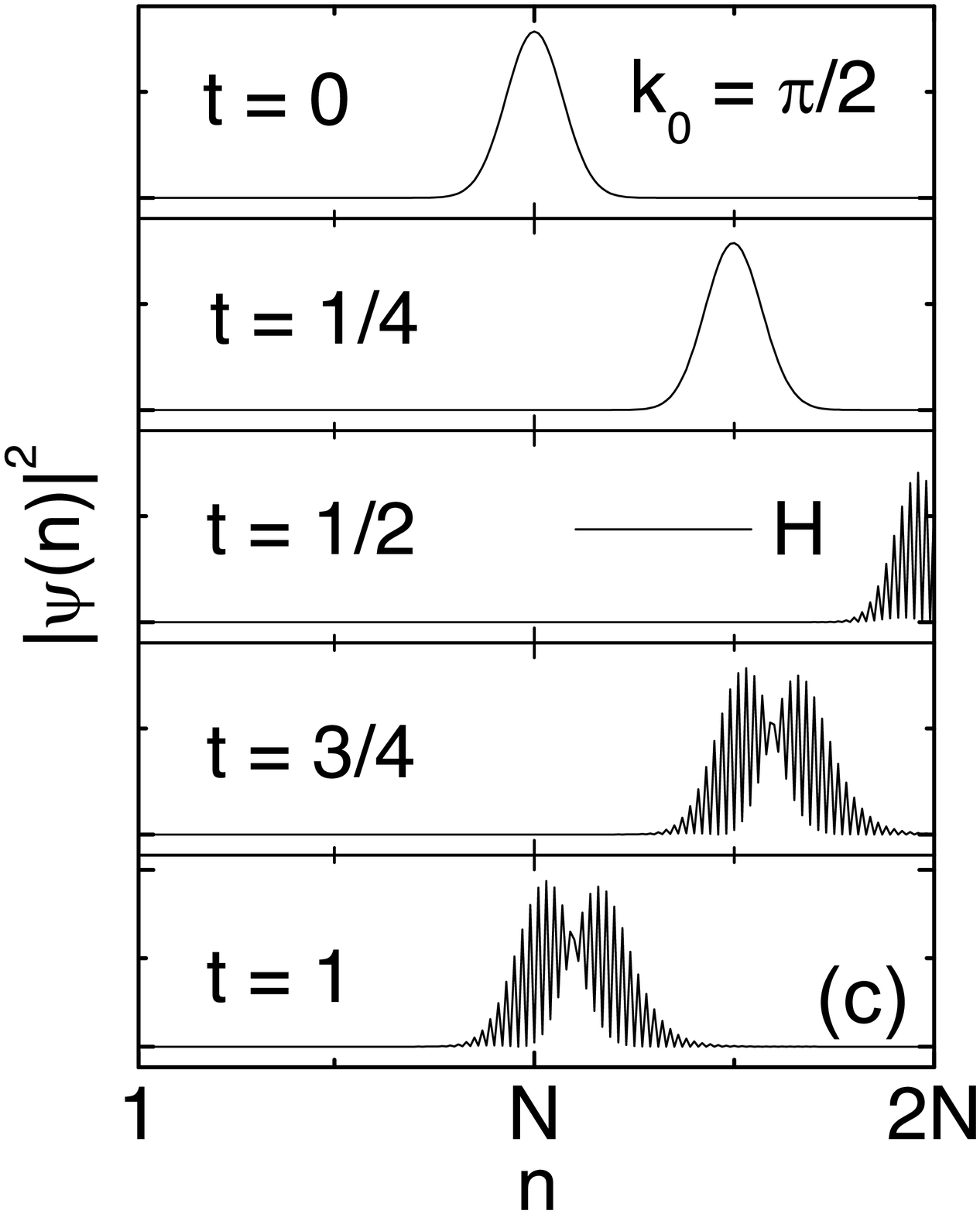}
\caption{The illustration of the time evolution of the same GWPs on
the same lattice as those in Fig.~\ref{fig.HC_Ex}, but under open
boundary conditions. We take $t$ in units of $T_{\mathrm{rev}}$ from
Eq.~(\ref{eq.T_rev}) in (a), and $T_{\mathrm{cir}}$ from
Eq.~(\ref{eq.T_cir}) in (b) and (c). One can find the dynamics of the
open boundary condition, which are the fractional revival and
non-spreading propagation, to be similar as those in the periodic
boundary conditions.}
\label{fig.HCC_Ex}
\end{figure*}

For $k_{0}=0$, at the instant $t$, we have
\begin{equation*}
\left\vert \Phi \left( 0,N_{\mathrm{A}},t\right) \right\rangle \propto
\sum_{k} e ^{-\frac{k^{2}}{8\alpha ^{2}}} e ^{- i \left(
N_{\mathrm{A}} + \frac{1}{2}\right) \frac{k}{2}} e ^{- i k^{2}
\frac{N^{2}}{ 2\pi T_{\mathrm{rev}}}t} \bar{\alpha}_{k} \left\vert 0
\right\rangle ,
\end{equation*}
where $T_{\mathrm{rev}}$ is the characteristic revival time that can
be estimated by \cite{Robinett} as
\begin{equation}
T_{\mathrm{rev}} = \frac{N^{2}}{\pi} \left\vert \left(
\frac{\partial^{2} \epsilon_{k}}{\partial k^{2}} \right)_{0}
\right\vert ^{-1} = \frac{2N^{2}}{\pi J_{\mathrm{e}}} .
\label{eq.T_rev}
\end{equation}
It shows that the fractional revival occurs due to the approximate
quadratic dispersion relation as if the wave packet evolves in the
Hamiltonian $h_{\mathrm{e}}$. For $k_{0}=\pi /2$, at the instant $t$,
we have
\begin{eqnarray*}
\left\vert \Phi \left(\frac{\pi }{2} , N_{\mathrm{A}} , t \right) \right\rangle &\propto & \sum_{k} e ^{- \frac{\left( k-\pi \right)
^{2}} {8\alpha ^{2}}} e ^{ - i N_{\mathrm{A}} \frac{k}{2}} \\
&& \times \left( e ^{- i v_{\frac{\pi}{2}}kt} \eta_{k}^{+}
\bar{\alpha}_{k} + e ^{ i v_{\frac{\pi }{2}}kt} \eta_{k}^{-}
\bar{\beta}_{k} \right) \left\vert 0\right\rangle ,
\end{eqnarray*}
where
\begin{equation}
v_{\frac{\pi}{2}} = \left\vert \left( \frac{\partial \epsilon _{k}}
{\partial k} \right)_{\pi} \right\vert = J_{\mathrm{e}},
\end{equation}
\begin{equation}
T_{\mathrm{cir}} = N\left\vert \left( \frac{\partial \epsilon_{k}}
{\partial k} \right)_{\pi} \right\vert ^{-1} =
\frac{N}{J_{\mathrm{e}}}, \label{eq.T_cir}
\end{equation}
which are the group velocity and circling period for a GWP of
$k_{0}=\pi /2$ in the effective ring.

To demonstrate the above-mentioned results, numerical simulations are
performed. We plot the illustration and the square of the Dirac norm
of the evolved GWPs of different $k_{0}$ in this
$\mathcal{PT}$-symmetric two-band ring as well as a uniform ring for
comparison in Fig.~\ref{fig.HC_Ex}. We should notice that at the
exceptional point, the gap disappears and these two bands merge. Under
this condition, the spectrum is the same as that of the effective
uniform ring with the hopping amplitude being $J_{\mathrm{e}}$. For
$k_{0}=0$, one can see that the GWP in the $\mathcal{PT}$-symmetric
ring has almost the same time evolution, which comes from the unequal
distribution of the initial state on the two bands in the momentum
space. For the wave packet with momentum $k_{0}=0$, it mainly locates
on the lower band of $\bar{\alpha}_{k}$ around $k \sim 0$ and rarely
locates on the upper band of $\bar{\beta}_{k}$, which satisfies the
quasi-Hermitian condition of $\left\vert g_{k} f_{k} \right\vert
\approx 0$. Under these circumstances, it can be treated as
quasi-Hermitian and the dynamics of the wave packet is similar as well
as in an effective uniform ring. And the Dirac probability of the wave
packet slightly deviates from unity, which is plotted in
Fig.~\ref{fig.HP_Ex}. The situation is similar for a $k_{0}=3\pi /8$
wave packet, the Dirac probability is also approximately conservative.
For $k_{0}=\pi /2$, although the wave packet consists of components
from both bands of $\bar{\alpha}_{k}$ and $\bar{\beta}_{k}$, the
quasi-Hermitian condition still fits. That is because, for the same
eigenvector $k$, one component from the two different bands is almost
zero and the other is finite while both zero on the broken states of
$\bar{\alpha}_{\pi }$ and $\bar{\beta}_{\pi }$. This meets the
quasi-Hermitian condition and hence the specific GWP exhibits an
analogous dynamical behavior as if in the effective Hermitian ring.

We should notice that the Hermiticity of the evolution on this
$\mathcal{PT}$-symmetric ring depends on not only the Hamiltonian, but
also the distribution of the wave packet on the two bands. At the
exceptional point, only two eigenstates are broken. When the
components of the GWP consist of neither the two states, the Dirac
norm will probably be quasi-Hermitian. The numerical simulations are
plotted in Figures~\ref{fig.HC_Ex} and \ref{fig.HP_Ex}. It shows that
the time evolution of $k_{0}=0 $ and $3\pi /8$ for the unbroken
Hamiltonian are about the same as those for the Hamiltonian near the
exceptional point. When the central momentum $k_{0}$ changes, the
distribution on the two bands changes (as plotted in
Fig.~\ref{fig.NHP_Ex}). The quasi-Hermitian condition of $\left\vert
g_{k} f_{k} \right\vert \approx 0$ is invalid, then the Dirac norm
deviates from unity apparently. In an unbroken ring with $\gamma
=0.19$, for the wave packet of $k_{0}=\pi /2$ and $-\pi /2$, which
contains the two unbroken eigenstates of $\alpha _{\pi }$ and $\beta
_{\pi }$ simultaneously, the quasi-Hermitian condition is no more
satisfied and the wave packet behaves in a non-Hermitian way as
plotted in Fig.~\ref{fig.NHP_NonEx}.

\section{Summary and discussion}
\label{sec.discus}

We have proposed a non-Hermitian $\mathcal{PT}$-symmetric two-band
model, which consists of dimerized hopping terms and staggered
imaginary on-site potentials. We have shown that such a model can have
real spectrum and exhibit Hermitian dynamical behavior, obeying
perfectly probability-preserving time evolution in terms of the Dirac
inner product. This fact indicates that the balanced gain and loss in
a non-Hermitian system can result in quasi-Hermiticity. Apparently,
such a dynamical behavior arises from the quasi-canonical commutation
relations in Eq.~(\ref{Canonical.CR}). The essence is the
translational symmetry of the model, which ensures the gain and loss
to distribute homogeneously. It is presumable that similar phenomenon
occur in a two-band chain system. It is more difficult to get the
analytical result when the open boundary is applied, compared to the
periodic boundary condition. In this case, numerical simulations have
be performed to compute the time evolution of a wave packet by direct
diagonalization of the Hamiltonian. We plot the numerical result for
the evolution of the same wave packet on the open chain in
Fig.~\ref{fig.HCC_Ex}. It shows that the open boundary condition does
not affect the obtained result so much. Since an open chain is much
more feasible to realize in practice compared to the ring, our results
can give a good prediction for the matter-wave dynamics in
experiments. The recent observation of the breaking of $\mathcal{PT}$
symmetry in coupled optical waveguides \cite{AGuo,Ruter} may pave the
way to demonstrate the result presented in this work.

\acknowledgments We acknowledge the support of the National Basic
Research Program (973 Program) of China under Grant No. 2012CB921900.

\end{document}